\documentclass[aps,twocolumn]{revtex4}
\usepackage{graphicx}
\usepackage{bm}
\usepackage{amsmath,amssymb}
\usepackage{times}
\begin{document}

\title{Heterogeneities in systems with quenched disorder}
\author{Mendeli H. Vainstein}
 \email{mendeli@iccmp.br}
 \affiliation{Instituto de F\'\i sica -- UnB  \\
CP 04455 \\
70919-970 Bras\'\i lia - DF - Brazil}

\author{Daniel A. Stariolo}
 \email{stariolo@if.ufrgs.br}
\homepage{http://www.if.ufrgs.br/~stariolo}
\altaffiliation{Associate researcher of the Abdus Salam International Centre for
Theoretical Physics.}
\affiliation{Instituto de F\'\i sica -- UFRGS \\ CP 15051 \\
91501-970 Porto Alegre RS - Brazil}

\author{Jeferson J. Arenzon}
 \email{arenzon@if.ufrgs.br}
\homepage{http://www.if.ufrgs.br/~arenzon}
\altaffiliation{Associate researcher of the Abdus Salam International Centre for
Theoretical Physics.}
\affiliation{Instituto de F\'\i sica -- UFRGS \\ CP 15051 \\
91501-970 Porto Alegre RS - Brazil}
\date{\today}

\begin{abstract}
We study the strong role played by structural (quenched) heterogeneities on 
static and dynamic properties of the Frustrated Ising Lattice Gas in two 
dimensions, already in the liquid phase. Differently
from the dynamical heterogeneities observed in other glass models in this case they 
may have infinite lifetime and be spatially pinned by the quenched disorder. 
We consider a measure of local frustration
show how it induces the appearance of spatial heterogeneities and how
this reflects in the observed behavior of equilibrium density distributions
and dynamic correlation functions.
\end{abstract}


\maketitle

\section{Introduction}

Much of the work in the theory of the glass transition has been
concentrated in the precursor phenomena present in the high 
temperature (or low density) phase on approaching the glass
transition.
The (equilibrium) dynamics in this region shows two qualitatively different regimes:
a short time relaxation associated with the rattling of particles inside
cages formed by their neighbors and a long time structural relaxation which, 
as density increases (or $T$ decreases), takes longer  and longer times
~\cite{Debenedetti96, DeSt01}. In supercooled liquids, the presence 
of a crossover temperature below which exponentially decaying correlation
functions typical of the liquid phase become stretched has been 
observed~\cite{SaDeSt98}. The origin of
this stretching of relaxations is associated to the gradual appearance of
heterogeneities in the space/time domain. Understanding the emergence
of these heterogeneities is at the heart of the nature of the glass
transition and has been the source of a lot of theoretical and experimental work
~\cite{Sillescu99,Ediger00,Richert02}. Nevertheless, a unifying microscopic theory
is still lacking and only a few universal properties related to heterogeneities have
been found. 

In real space, dynamical heterogeneities correspond to particles with different
mobilities. In computer simulations of simple glass formers 
the presence of {\em fast} and {\em slow} particles has been observed.
The typical time scales of the two types of particles
 can be identified for example by examining the behavior of
the non-gaussian parameter associated with the van Hove correlation function
~\cite{KoDoPlPoGl97,DoDoKoPlPoGl98}. This leads to the identification of a
growing time scale which seems to diverge at the mode coupling transition
temperature ($T_{MCT}$). This characteristic time arises naturally when computing a
suitable time dependent four point correlation function or {\em dynamical
non-linear susceptibility}~\cite{GlNoSc00,DoFrPaGl02}. As the temperature 
approaches $T_{MCT}$ from above, this dynamical response shows a growing
peak at a characteristic time $t^*$ and then relaxes at long times to its
equilibrium value. This growing characteristic time of the four point functions
has been associated with the presence of long lived dynamical heterogeneities
(clusters of very slow or immobile particles).

From a landscape point of view, the emergence of stretching  seems to be 
consequence of the gradual confinement
of the system in low (free) energy regions of phase space surrounded by growing
barriers as temperature decreases. The confinement is due in part to the
dramatic decrease of escape directions in configuration space on
approaching the glass transition~\cite{BrBhCaZiGi00,AnDLRuScSc00,GrCaGiPa02}. In
simple models of binary mixtures it has been found that stretched relaxations
are associated with long lived superstructures or ``metabasins'' in configuration
space~\cite{BuHe00}. The relaxation from single metabasins shows a stretched behavior
and different metabasins relax with different characteristic times, giving rise
to the overall stretching observed for example in the $\alpha$ region of the 
incoherent scattering function.

At variance with the enormous amount of work on the problem of heterogeneities
in glasses, much less is known about them in systems with quenched disorder
~\cite{PoCoGlJa97,GlJaLoMaPo98,BaZe99}.
Differently from true glasses, where the inhomogeneities are
dynamically created by the self induced local frustration and
persist during a given timescale, the quenched disorder
induces structural heterogeneities that may have infinite lifetime. In this
sense they can be called {\em quenched heterogeneities}. Since
the inhomogeneities are pinned by the disorder, they may be
easier to detect and characterize than in systems with self induced disorder.
Thus, the posed question is to what extent these heterogeneities influence
the relaxation properties and how different regions of the sample differ
from each other. Moreover, how do they affect the different degrees of 
freedom present in the system (like particles and spins)? 

To answer these questions, in this work we address the problem 
of heterogeneities in systems with
quenched disorder by focusing the analysis on the behavior of a two
dimensional version of the Frustrated
Ising Lattice Gas (FILG) which has been proposed as a simple lattice
glass model~\cite{NiCo97}. Nevertheless it is
known that the presence of quenched disorder in the definition of
the model introduces some features not typical of real glasses. In
particular it acts as a pinning field for heterogeneities, some of
which can have infinite lifetime. This is reflected in the behavior
of dynamical correlation functions and in the non-linear susceptibility
and compressibility as will be shown below. In this context we will 
consider a measure of local frustration~\cite{PoCoGlJa97,GlJaLoMaPo98},
show how it induces the appearance of spatial heterogeneities and how
this reflects in the observed behavior of equilibrium density distributions
and dynamic correlation functions. For the sake of comparison with other
glassy systems without quenched disorder we also discuss the behavior of
the non-linear susceptibility associated with the spin degrees of freedom 
and of the non-linear compressibility of the density variables. Finally, an
analysis of the universal behavior in terms of hole variables recently
introduced~\cite{LaReMcGrTaDa02,DaLaGrMcZaTa02} shows that this system 
seems to be in a different universality class
than the previously investigated glass models without disorder.

\section{The Model}

The Frustrated Ising Lattice Gas~\cite{NiCo97,FiCaCo00,StAr99,ArRiSt00} is
defined by the Hamiltonian
\begin{equation}
H = -J \sum_{<ij>} (\varepsilon_{ij} \sigma_i \sigma_j - 1) n_i n_j -
\mu \sum_i n_i\ .
\end{equation}
At each site of the lattice there are two different dynamical
variables: local density (occupation) variables $n_i=0,1$ ($i=1 \ldots
N$) and internal degrees of freedom, $\sigma_i=\pm1$. The usually
complex spatial structure of the molecules of glass forming liquids,
which can assume several spatial orientations, is in part responsible
for the geometric constraints on their mobility. Here we consider the
simplest case of two possible orientations, and the steric effects
imposed on a particle by its neighbors are felt as restrictions on its
orientation due to the quenched variables $\varepsilon_{ij}$ connecting
nearest neighbors sites. 
The first term of the Hamiltonian ensures that when $J$ is large, any 
two neighboring particles will have their spins satisfying the connection 
between them. Finally $\mu$ represents a chemical
potential ruling the system density (at fixed volume).

This model has been studied mainly through 
simulations~\cite{NiCo97,FiCaCo00,ArRiSt00} in $3d$ and analytically
in mean field~\cite{ArNiSe96,CrLe02}. In the low density (high temperature)
phase, the system behavior is liquid like, the dynamics is fast,
time-translationally invariant and obeys the fluctuation-dissipation
theorem. By increasing the density (or lowering the temperature), the system 
has a spin glass transition characterized by
a divergence of the static nonlinear susceptibility. This transition point
is located quite close to the point where the system suffers a dynamical
arrest, where the particles are no longer able to diffuse. After a 
sub-critical quench,
the dynamics is slow and history dependent. Besides that, it presents
 aging and violates the
fluctuation-dissipation theorem (we refer the interested reader
to \cite{ArRiSt00} and references therein).

\section{Static properties}

Since bonds are created at random (with probability $1/2$ of being
$\pm 1$), the system is not homogeneous at the lenghtscale of the
lattice spacing. Each site belongs to several minimal plaquets and
there is a probability that all (or neither) of these 
are frustrated. This would influence the dynamics at a local
level since the system is no longer invariant under spatial
translations. Thus we may ask which are the consequences of these 
quenched heterogeneities on the equilibrium properties of the system.

When  evaluating the partition function
at finite density, the configurations in which particles close any
frustrated loop will have zero probability. The average density will no
longer be site independent and the density distribution $P(\rho)$
will be inhomogeneous, i.e. will show spatial fluctuations. Homogeneity 
will only be recovered when $\rho\to 0$. For small densities, the 
distribution will be gaussian while deviations
from gaussianity should be evident as the density increases.
The forbidden configurations introduce
spatial inhomogeneities that are long lived as they derive from
the quenched {\it random} underlying connections. In other lattice models,
as well as in structural glasses, the disorder is not quenched, 
what restores translational invariance if the measure time is
much larger than the dynamical heterogeneities characteristic time. 
An example of a system with quenched but not random frustration,
that does not present these inhomogeneities is the fully
frustrated version of the FILG.

\begin{figure}[h]
\includegraphics[width=4cm]{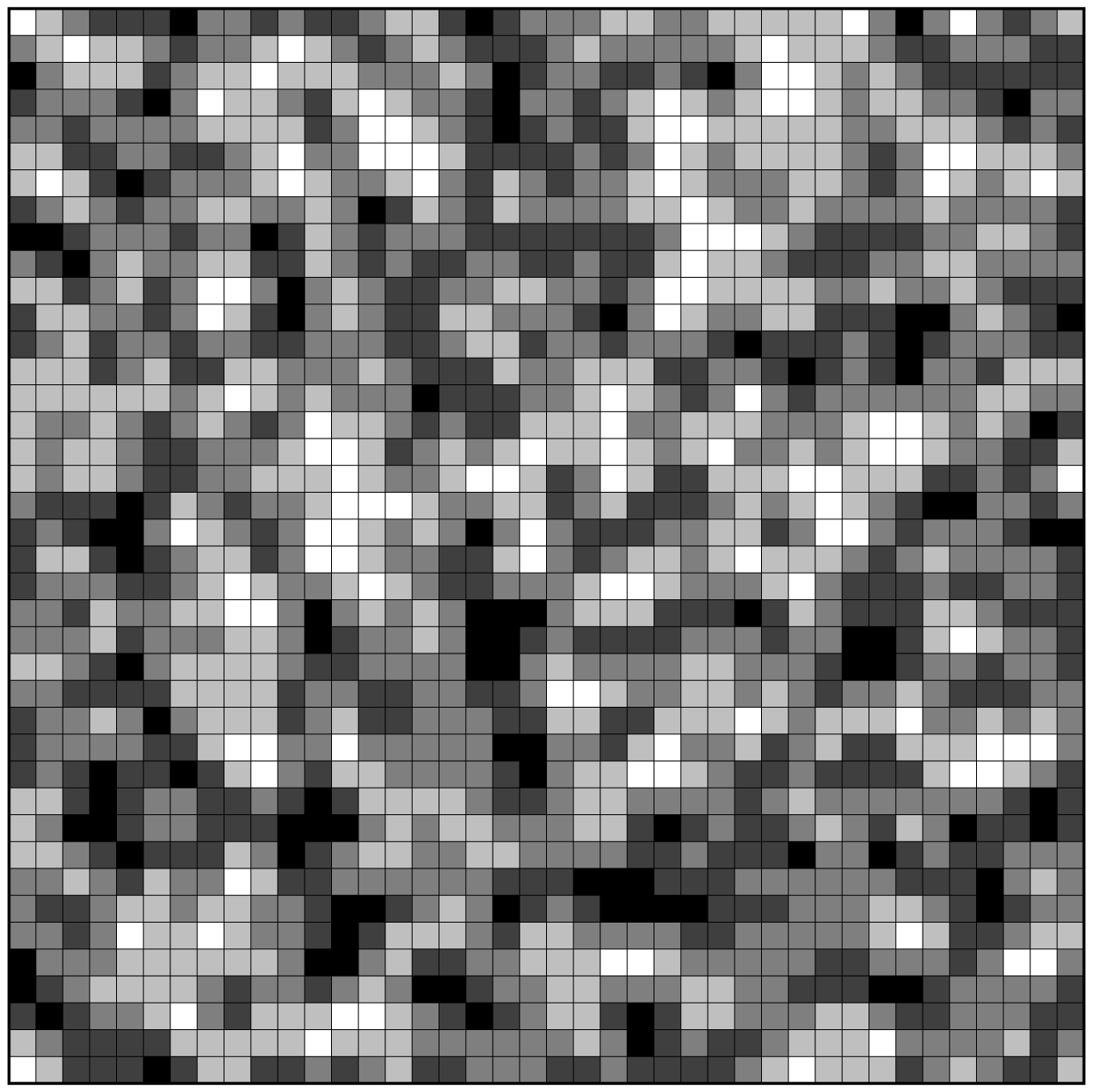}
\includegraphics[width=4cm]{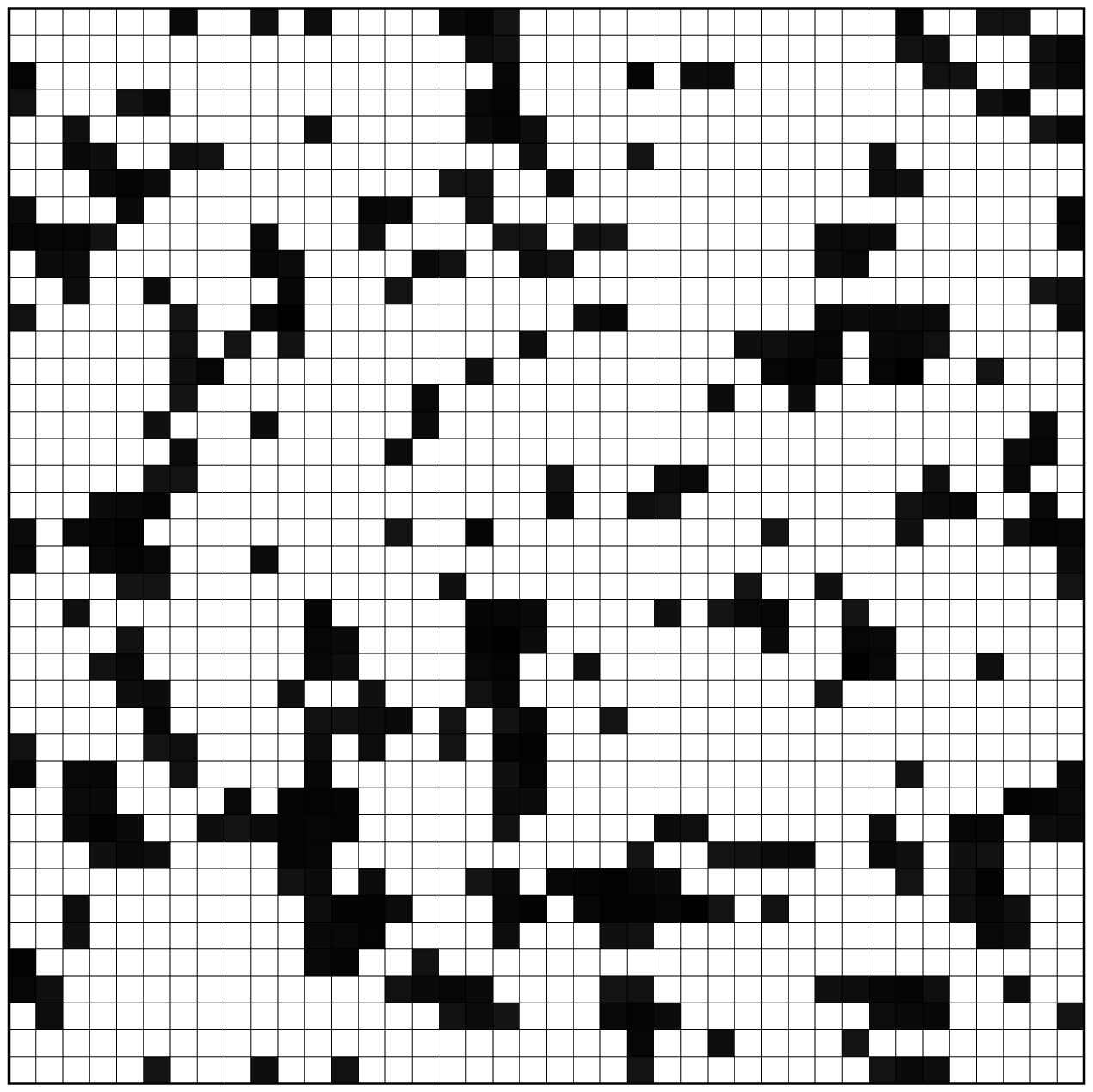}
\caption{$a)$ Example of a $\{\gamma_i\}$ configuration. Black sites are 
those whose four neighbor plaquets are
non frustrated ($\gamma_i=4$) while white sites have all frustrated
($\gamma_i=-4$). Intermediate values of $\gamma_i$ are represented
by different degrees of grey. $b$) Sites whose average density is higher
than 0.9 for $\mu=4$, corresponding to a global $\rho\simeq 0.74$.}
\label{fig.plaq} 
\end{figure}

To quantify this we use the (local) coarse-grained quantity
\begin{equation}
\gamma_i\equiv \sum_{{\cal P}(i)} \prod_{j\in {\cal P}(i)} \varepsilon_{ij}
\end{equation}
where ${\cal P}(i)$ are the minimal plaquets containing the site
$i$. That is, for every site, we consider the 4 minimal plaquets
that contain that site (12 in $d=3$) and we check how many of these plaquets
are frustrated, i.e. the product of the connections in the loop is negative. 
This local frustration parameter was originally
considered in the context of dynamical heterogeneities
 in refs.~\cite{PoCoGlJa97,GlJaLoMaPo98}.
In this nearest neighbor plaquet approximation
sites that are maximally frustrated have all of its 4 plaquets
with negative products and sites with no frustration have all plaquets
with positive products. In other words, $\gamma_i$ may assume 
values between $-4$ and 4. Figure~\ref{fig.plaq}$a$ shows
a configuration with inhomogeneous distribution of local
frustration. 

\begin{figure}[h]
\includegraphics[width=5cm,angle=270]{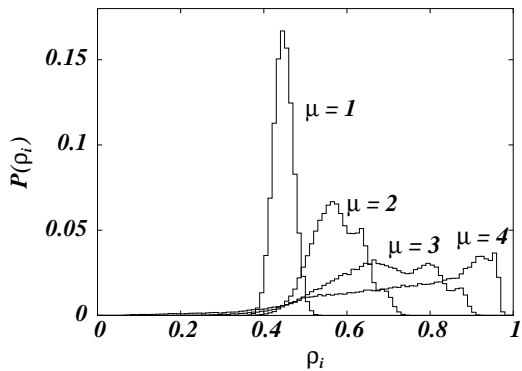}
\caption{\label{fig.prho} Distribution of local densities for several
values of $\mu$ and $L=20$ in $2d$. Notice the multi-peak structure
that becomes more apparent as $\mu$ increases.}
\end{figure}

We now consider how these structural clumps change the
system's behavior. In fig.~\ref{fig.plaq}$b$ we show, for the same realization
of disorder of fig.~\ref{fig.plaq}$a$, only the sites that have an 
averaged occupation greater than 0.9. In this figure, $\mu=4$, and
 the total average density is roughly 0.74, well below 0.9. The
presence of preferred positions is a direct consequence of the existence
of islands of low frustration, as can be seen by comparing
figures \ref{fig.plaq}$a$ and \ref{fig.plaq}$b$. As the chemical potential
increases, the number of these sites also increase, as expected. 
These heterogeneities make the system not translationally invariant
even in the low density liquid phase. This can be made clear by
measuring the distribution $P(\rho_i)$ of local averaged densities 
$\rho_i=\left\langle n_i\right\rangle$ as shown in fig.~\ref{fig.prho}. In homogeneous
systems all sites have the same average density and the 
distribution, for a finite number of measurements, is a gaussian centered in 
this value. Examples might be a non disordered version of the model
considered here (with fully frustrated connections) or models for 
structural glasses without quenched variables. In the last case, 
the annealed character of 
the frustration allows the homogeneity to be recovered again.
The broadness of the curves does not seem related neither to finite
average times nor to finite lattice sizes, but rather to the
fact that, when not limited to the nearest neighbor plaquet
approximation, there are several values that the local frustration
may assume generating in this way the broad distributions seen
in the above figures.

\begin{figure}[h]
\includegraphics[width=5cm,angle=270]{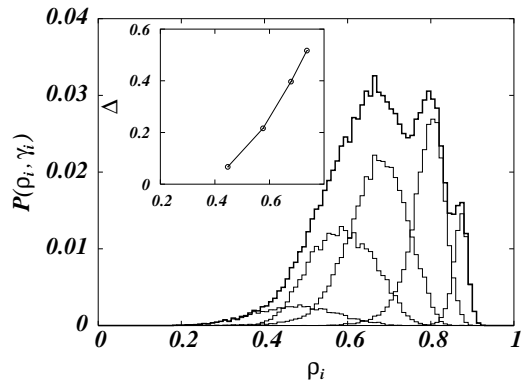}
\caption{\label{fig.prhosep} Distribution of local densities for $\mu=3$ and 
$L=20$ in $2d$. The bold curve is the sum of the several others, $P(\rho_i)$. 
From left to right, we have 4, 3,\ldots,0 locally frustrated loops. The
distribution has an internal structure. Inset: The distance $\Delta$ between
the average densities for extreme values of $\gamma$, $P(\rho_i,4)$ and
$P(\rho_i,-4)$.}
\end{figure}

To make clear
the origin of this broad distribution, we considered the density
distribution $P(\rho_i,\gamma_i)$ separately for each value of the local 
frustration $\gamma_i$, which has a well defined density around which 
its particular distribution is centered. The
result is seen in fig.~\ref{fig.prhosep}. Moreover, the separation
between these curves increases as the density increases and the
deviation from gaussianity becomes apparent.
We may define $\Delta$ as the difference between the average
densities of the peaks associated with the lowest and highest
values of $\gamma$, $\Delta\equiv \left\langle \rho_i\right\rangle_4
- \left\langle \rho_i\right\rangle_{-4}$ where the averages 
$\left\langle ... \right\rangle_{\gamma_i}$ are over $P(\rho_i,\gamma_i)$. 
When there is no difference, $\Delta=0$, the
particles do not feel the underlying landscape and all sites 
have the same average equilibrium density. In this model this happens for very low
densities. As the density increases, the dynamics becomes
landscape influenced and $P(\rho_i)$ has a more complex structure,
no longer being a simple gaussian. This is exemplified in
fig.~\ref{fig.prhosep} for $\mu=3$. One can also note that
the distributions are asymmetric around $\gamma_i=0$: for example,
in the case of $P(\rho_i,-2)$ and $P(\rho_i,2)$ the former is
broader and shorter while the latter is higher  and more
concentrated. A possible explanation is that sites with negative values
of $\gamma_i$ are more influenced by higher order plaquets than
sites with positive $\gamma_i$, thus having a broader distribution.

\begin{figure}[h]
\includegraphics[width=5cm,angle=270]{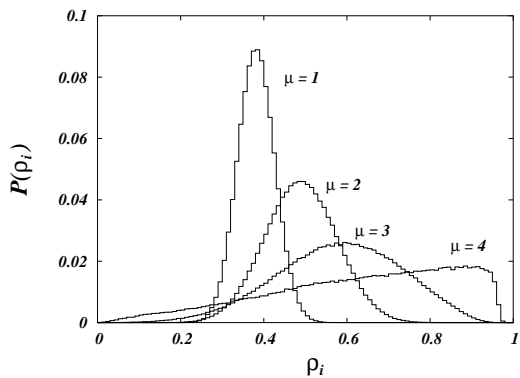}
\caption{\label{fig.prho3} Distribution of local densities for several
values of $\mu$ and $L=10$ in $3d$. 
Differently from the $2d$ case, the curves do not seem to present
internal structure.}
\label{histo.3d}
\end{figure}

In $d=3$ the overall picture is similar to that of $d=2$. Nevertheless some
differences are evident when comparing figures \ref{fig.prho} and \ref{histo.3d}
and one may wonder if the minimum plaquet approximation used in the definition
of the local frustration parameter works in $d=3$ as it does in $d=2$.
Still within the same approximation, the distribution of local densities for
different local frustrations presents also different peaks (not shown),
a manifestation of the lack of translational invariance also in $d=3$. 

\begin{figure}[h]
\includegraphics[width=5cm,angle=270]{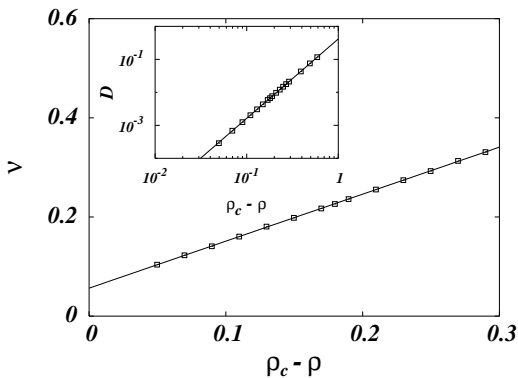}
\caption{The fraction of holes $\nu$ as a function of $\rho-\rho_c$
for $L=50$. Notice that $\nu$ does not vanish at $\rho_c\simeq 0.79$. The behavior
near $\rho_c$ is linear with the density. Inset: diffusion
coefficient as a function of $\rho_c-\rho$. The diffusivity goes
to zero as $(\rho_c-\rho)^{\phi}$ with $\phi\simeq 2.4$.}
\label{fig.holes}
\end{figure}

Recently, a description of the jamming transition has been
introduced that unveiled some universal mechanism leading to
the dynamical arrest that happens at $\rho_c$. The notion of hole
is introduced: empty sites that have at least
one neighbor particle able, due to the energetic or to
the kinetic constraints, to jump to the initial empty site.
We measure, for several {\em fixed} densities~\footnote{In this
case, when diffusing, the particle carries its spin.},
 the density of holes,
$\nu$. It has been conjectured, and supported by numerical
simulations on some lattice models~\cite{LaReMcGrTaDa02,DaLaGrMcZaTa02}, 
that the diffusivity 
depends on $\nu$ as $D\sim (\nu-\nu_0)^2$ where $\nu_0$ is a
possible residual density of holes at $\rho_c$ (rattlers) that do not
contribute to the diffusion. We show our results for the FILG in $d=2$
in fig. ~\ref{fig.holes}. We obtain $D\sim (\nu-0.056)^{2.4}$.
The Kob-Andersen
is another example of model where $\nu_0\neq 0$. Some things are
remarkable: first, we notice that the FILG belongs
to a different universality class. Second, the value of $\nu$ at
$\rho_c$ is much larger than in other models.
Notice also that the behavior of $\nu$ near $\rho_c$ is linear.
The fact that the FILG does not belong to the same universality
class of several other models is very interesting and the reason
may be related to the inhomogeneity of the underlying
connections structure. Indeed, we measure the density of holes
accordingly to the $\gamma_i$ of the hole site. For small
values of $\rho$ there is no difference between these values
and all curves collapse. The fact that for higher 
densities these curves fall apart is an indication that the
holes also have preferred sites and tend to form small
clusters. Thus, near $\rho_c$, the mechanism leading to
diffusion is no longer the pairwise collisions between
holes but a more complicated process involving these small
clusters, what may lead to a failure of the $\nu$-square 
law~\footnote{Deviations from this law were also observed in the
KA model in the density regime near the dynamical arrest
(M. Sellitto, private communication.)}.
Further work along this line has to be done in order to
clarify this issue.

\section{Dynamical properties}

\begin{figure}[h]
\begin{center}
\includegraphics[width=6cm,angle=270]{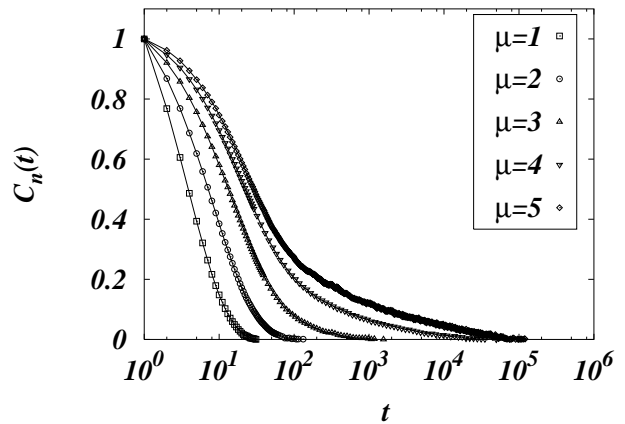}
\end{center}
\caption{Equilibrium particle correlations $C_n(t)$ for several
values of $\mu$ for a dynamics with spin flip, particle creation-destruction
and diffusion.} 
\label{fig.cn}
\end{figure}

In Ref.\cite{StAr99} the nonequilibrium behavior of the model was
shown to present interesting aging properties similar to those
observed in glass forming systems. Here we address 
equilibrium density correlations instead and the role of
heterogeneities in the low density phase. The density autocorrelations 
are defined by
\begin{equation}
c_n(t) = \frac{1}{N}\sum_i \left[ 
\left\langle n_i(t)n_i(0)\right\rangle 
      - \left\langle n_i(0)\right\rangle^2 
\right]
\end{equation}
and $C_n(t)\equiv c_n(t)/c_n(0)$,
where the averages are both over samples and initial states. The 
densities are also averaged over the thermal history in each
sample.
Notice that only in the case where the system is homogeneous,
that is, the average quantities are invariant under spatial translations,
we can substitute the last term in the numerator by $\rho^2$,
otherwise $\sum_i\left\langle n_i\right\rangle^2\neq 
\left\langle\sum_in_i\right\rangle^2$. In the region of densities
considered here, these correlations do not present the usual two step dynamics 
observed in other models as can be seen in fig.~\ref{fig.cn}. Nevertheless
two regimes can be clearly observed and all the curves can be fitted by
stretched exponentials  (not shown) in the first regime. The second regime at longer
times observed for the largest densities can also be fitted by stretched
exponentials with other parameters and probably will tend to develop a
plateau at higher densities.
 
\begin{figure}[h]
\begin{center}
\includegraphics[width=6cm,angle=270]{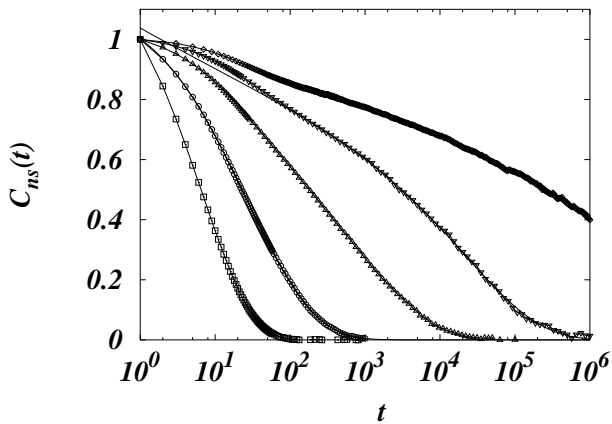}
\end{center}
\caption{Equilibrium diluted spin correlations $C_{ns}(t)$ for several
values of $\mu$ for a dynamics with spin flip, particle creation-destruction
and diffusion. The symbols are the same as in fig.~\ref{fig.cn}.} 
\label{fig.cns}
\end{figure}

\begin{figure}[h]
\begin{center}
\includegraphics[width=6cm,angle=270]{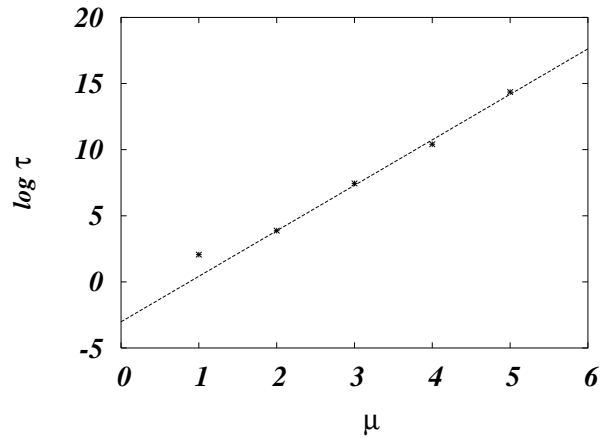}
\end{center}
\caption{Relaxation times of the diluted spin correlations as a function of
the chemical potential in a semi-log plot.} 
\label{fig.taus}
\end{figure}

For the same values of the chemical potential or mean density shown in
fig.~\ref{fig.cn}, the diluted
spin correlations show a different behavior. The correlations
between spins on occupied sites are defined as
\begin{equation}
C_{ns}(t)=\frac{1}{N}\sum_i \left\langle
n_i(t)S_i(t) n_i(0)S_i(0) \right\rangle\;.
\end{equation}
The results for several values of $\mu$ are shown in
fig.~\ref{fig.cns} together with the best fits to stretched exponential
decays for the long time regime. 
Interestingly, the correlations 
$C_n$ and $C_{ns}$ decay with
very different timescales: density correlations decay much
faster than the diluted spin correlations. This can be understood as
follows. Particles can be created/destroyed and diffuse through the
lattice, forming clusters of neighboring particles. Once they
belong to the same cluster, the nature of the interaction enforces
them to have spins satisfying all the bonds in the cluster, otherwise
they should move apart. Thus, when a site contributes for the correlation
at different times, the contribution will be positive, unless there is
enough time for the whole cluster to flip all its spins. Moreover, the
percolation transition is just below $\mu=3$ and the considered clusters
are large, making the decorrelation time quite large.
The diluted spin correlations start to develop an incipient plateau as 
the value of the chemical potential increases.
We expect that by further increasing the value of $\mu$, this
plateau will be more noticeable and the characteristic decay time will
tend to diverge at the dynamical transition. In fig.~\ref{fig.taus} we
show an Arrhenius plot of the relaxation times showing that they tend
to diverge for $\mu \to \infty$. This is different from the behavior 
observed in the three dimensional model~\cite{CaCo01}, which shows a power
law divergence at a finite value of $\mu$. Thus, while the three dimensional
model presents a spin glass transition at a finite chemical potential
the two dimensional model only has a transition at $\mu \to \infty$. 

\begin{figure}[h]
\begin{center}
\includegraphics[width=6cm,angle=270]{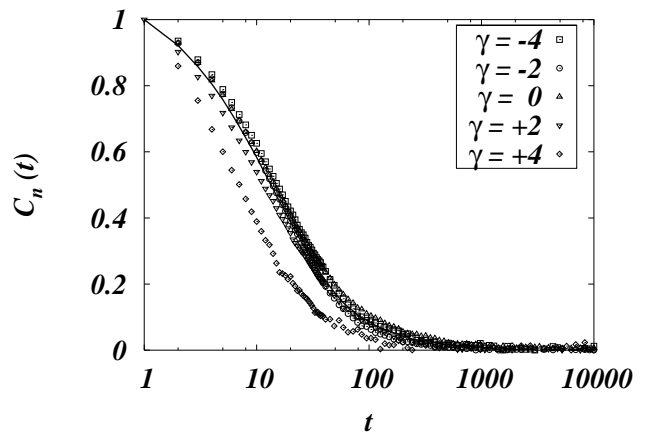}
\end{center}
\caption{Equilibrium particle correlations $C_{n}(t)$ for
the five possible values of $\gamma_i$ and $\mu=3$. The line is the
full correlation, from fig.~\ref{fig.cn}.} 
\label{fig.cn_mu03}
\end{figure}

\begin{figure}[h]
\begin{center}
\includegraphics[width=6cm,angle=270]{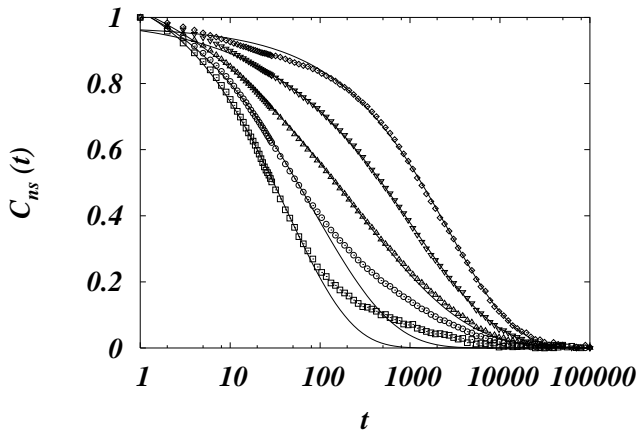}
\end{center}
\caption{Equilibrium dilute spin correlations $C_{ns}(t)$ for
the five possible values of $\gamma_i$ and $\mu=3$. The symbols
are the same as in fig.~\ref{fig.cn_mu03}. Notice that low
frustration sites already show a small plateau, although in the
full correlation (fig.~\ref{fig.cns}), this is not yet noticeable.} 
\label{fig.cns_mu03}
\end{figure}

Both correlations can also be measured only for sites presenting
a given value of the frustration parameter $\gamma_i$. 
The curves for the 5 possible values
(in $d=2$) are shown in figs.~\ref{fig.cn_mu03} and \ref{fig.cns_mu03}
for $\mu=3$. Fits to stretched exponential decay are also shown for the
diluted spin correlations.
They are normalized to unity but, recalling fig.~\ref{fig.prhosep}, the number
of sites contributing to each one of the curves is different. The behavior for
$C_n(t)$ is very different and almost all curves give the same correlation, 
only sites with $\gamma_i=4$
differ slightly from the others (and their contribution to the overall
curve is small). On the other hand, for $C_{ns}(t)$
the curves are strongly resolved and we notice that the dynamics
is faster in those sites in highly frustrated regions. This may be
due to the fact that since particle creation-destruction is allowed,
particles appearing in these regions have a higher probability to
go elsewhere or be destroyed (as these are regions of less than average
densities), what makes the decorrelation time
small. On the other hand, low frustration regions are those preferred
by the particles and correlations take much more time to decay (as these
are regions of larger than average densities).
The total correlation (not shown) follows closely the $\gamma_i=0$
or mean frustration curve.
Note that all local correlations show stretched exponential relaxations
irrespective of the value of the local frustration being high or low. In
this scenario the stretched relaxation does not emerge as a consequence
of some kind of convolution of local exponential relaxations with
different time scales. At this level of description, the stretching, 
although heterogeneous due to different
local frustration, is intrinsic. As the density increases this scenario does not 
change qualitatively.

It is interesting to remark that for high values of $\mu$ the
decay of the diluted spin correlation starts to develop a plateau and the
sites responsible for this behavior are those that are less
frustrated. The effect is not much pronounced because the
summed up effect of $\gamma_i=-2$ and 0 sites is dominant.

\begin{figure}[h]
\begin{center}
\includegraphics[width=6cm,angle=270]{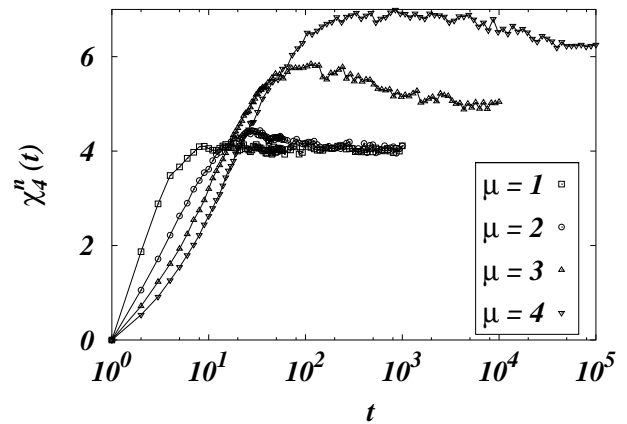}
\end{center}
\caption{$\chi_4^n(t)$ as a function of time for different
chemical potentials. For increasing chemical potential (hence
densities) the peak increases and shifts to longer times.} 
\label{fig.het}
\end{figure}

As the dynamical heterogeneities have attracted much attention
recently, several quantities have been devised to quantify 
them~\cite{Sillescu99}.
Among them, of particular interest is the four point correlation
(or dynamical nonlinear response) defined as
\begin{equation}
\chi_4^n(t)=N \left( \left\langle C_n^2(t) \right\rangle - 
\left\langle C_n(t) \right\rangle^2 \right)\;.
\label{eq.x4}
\end{equation}
Away from the transition, this function presents a maximum
at a time $t^*$ that is a measure of the timescale during which
the particles forming the heterogeneity are correlated. The
long time behavior of this quantity is a measure of its 
equilibrium value. 
In figure~\ref{fig.het} we plot $\chi_4^n(t)$.
Analogously to what happens in other glass models it presents
a broad peak that shifts to longer times and gets higher as the
density increases. Nevertheless an important difference from other models is
that after the peak the non-linear compressibility does not decay too
much and develops a density-dependent plateau at long times. The
persistence of density correlations is a clear signature of the
presence of very long lived heterogeneities, a consequence of the
pinning effect of the quenched disorder present in the model.
The possible divergence of the equilibrium (infinite time) limit of
the non-linear compressibility should point to the existence of a
thermodynamic transition associated with the density degrees of
freedom not yet observed in this model.

\begin{figure}[h]
\begin{center}
\includegraphics[width=6cm,angle=270]{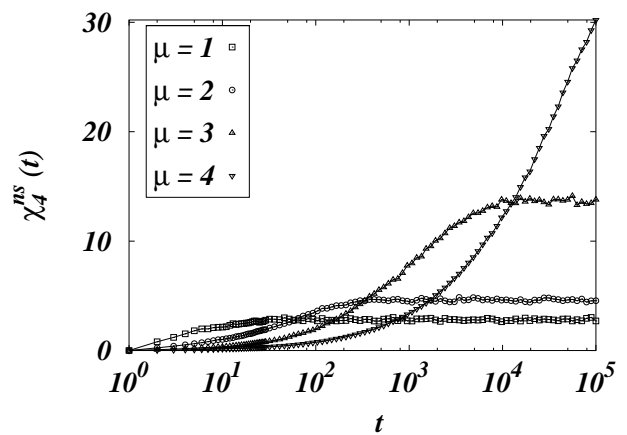}
\end{center}
\caption{$\chi_4^{ns}(t)$  as a function of time for different
chemical potentials. Note here the saturation after a crossover
time and the stronger growth of the subsequent plateau at variance
with the behavior of $\chi_4^{n}(t)$ in fig. ~\ref{fig.het}.}
\label{fig.hetns}
\end{figure}

Analogously, for diluted spin correlations we can evaluate eq.~\ref{eq.x4}
using $C_{ns}(t)$ in the place of $C_n(t)$. The resultant $X_4^{ns}(t)$
is seen in fig.~\ref{fig.hetns}. The behavior is similar to that observed
for the density variables, but instead of a peak we observe a crossover time after 
which this non-linear susceptibility saturates in a finite density dependent
value. In this case the height of the plateau seems to diverge much more
rapidly than for the compressibility giving more plausibility to the presence
of a thermodynamic transition in the spin variables, probably for $\mu \to \infty$.
The crossover to a plateau, instead of the presence of a peak, clearly points to
the appearance of a persistent correlation length associated with the spin degrees
of freedom. It would be interesting to check if these persistence
is a manifestation of a growing structural correlation length which
might diverge at an equilibrium spin glass transition similar to that present in
d=3.

\section{Conclusions}

In this paper we considered the role played by heterogeneities
present in a model with quenched disorder. Differently from
self-induced frustration models here the heterogeneities may be
pinned by the local frustration and this allows a simple
characterization of them. The effects of these pinned heterogeneities on
time correlation functions and distribution of local densities have been analyzed.
The system forms clumps in 
the liquid phase even when the particles either do not interact or
strongly repel each other~\cite{KlGoRaClMe94}.
Moreover, these clumps are long-lived and it would be interesting to study
their size distribution and how they are correlated~\cite{JoMeGoKlMo98}.

A simple measure of local frustration~\cite{PoCoGlJa97} was considered and a strong correlation
with the inhomogeneous density distribution was found. The use of $\gamma_i$
allowed to decompose the local density distributions in components related to
the degree of local frustration. However,
the use of $\gamma_i$ as a measure of the local frustration considering 
only the minimal plaquets has to be considered as a first order
approximation. Indeed, higher order plaquets, involving larger loops,
have to be taken into account in order to obtain the full dependence
of $\rho_i$ on $\gamma_i$. For example, in
the $d=2$ case, there exists 12 plaquets of second order (those having
6 bonds). It is an open problem to determine to what extent the present
results may change by considering the effects of larger loops in
$\gamma_i$ and how they determine the average local density
and affect the relaxation properties. 
It was shown also that dynamic correlations can be decomposed accordingly
to the local values of frustration and in each case the relaxations are
essentially stretched although the degree of stretching is of course dependent
on the degree of local frustration.
It would be interesting to study the possible relation between {\em local} correlation 
functions in the out-of-equilibrium regime and
the violation of the fluctuation-dissipation 
theorem~\cite{BaZe99,CaChCuKe02,MoRi03,Berthier03b} with the local frustration as defined by
the parameter $\gamma_i$.

Further insight into the relation of the FILG with other models of glasses and spin
glasses can be gained by analyzing non-linear responses. We showed that two
such responses associated with different degrees of freedom in the model behave
differently in some important aspects. At variance with models of glasses,
the spin non-linear susceptibility shows a crossover time after which the
response saturates instead of the characteristic peak observed for example in
models with constrained dynamics. Also the rapid growth of the plateau at long
times suggests the developing of a structural correlation length associated
with a possible spin glass transition. This remains to be confirmed.
The non-linear compressibility instead
shows a broad peak but with a modest further relaxation to a density dependent
plateau. In this case one can associate this plateau with the formation of 
very long lived {\em quenched heterogeneities}. From the observed behavior of
these non-linear responses one can conclude that the FILG does not behave
neither as a proper spin glass nor as a glass sharing some characteristics 
with both kinds of systems.

Although the model considered here has been originally introduced as an 
attempt to obtain a finite dimensional lattice system with glassy properties
similar to structural glasses the main drawback is the lack of invariance under 
spatial translations already in the liquid phase. There are several ways to
restore this invariance. For example, by considering spatially 
coarse-grained variables or by including a temporal evolution of the
bonds~\cite{FiCaCo00}. As it stands this model still may be
interesting to study transport in disordered media 
(see~\cite{OsBeBuMo02} and references therein) which
has been of much interest recently.

\begin{acknowledgments}
We acknowledge interesting conversations with
Annalisa Fierro and Yan Levin. 
Work partially supported by the brazilian
agencies FAPERGS and CNPq.
\end{acknowledgments}


\begin{thebibliography}{32}
\expandafter\ifx\csname natexlab\endcsname\relax\def\natexlab#1{#1}\fi
\expandafter\ifx\csname bibnamefont\endcsname\relax
  \def\bibnamefont#1{#1}\fi
\expandafter\ifx\csname bibfnamefont\endcsname\relax
  \def\bibfnamefont#1{#1}\fi
\expandafter\ifx\csname citenamefont\endcsname\relax
  \def\citenamefont#1{#1}\fi
\expandafter\ifx\csname url\endcsname\relax
  \def\url#1{\texttt{#1}}\fi
\expandafter\ifx\csname urlprefix\endcsname\relax\def\urlprefix{URL }\fi
\providecommand{\bibinfo}[2]{#2}
\providecommand{\eprint}[2][]{\url{#2}}

\bibitem[{\citenamefont{{Debenedetti}}(1996)}]{Debenedetti96}
\bibinfo{author}{\bibfnamefont{P.}~\bibnamefont{{Debenedetti}}},
  \emph{\bibinfo{title}{Metastable liquids, Concepts and Principles}}
  (\bibinfo{publisher}{Princeton University Press},
  \bibinfo{address}{Princeton}, \bibinfo{year}{1996}).

\bibitem[{\citenamefont{Debenedetti and Stillinger}(2001)}]{DeSt01}
\bibinfo{author}{\bibfnamefont{P.~G.} \bibnamefont{Debenedetti}}
  \bibnamefont{and} \bibinfo{author}{\bibfnamefont{F.~H.}
  \bibnamefont{Stillinger}}, \bibinfo{journal}{Nature}
  \textbf{\bibinfo{volume}{410}}, \bibinfo{pages}{259} (\bibinfo{year}{2001}).

\bibitem[{\citenamefont{{Sastry} et~al.}(1998)\citenamefont{{Sastry},
  {Debenedetti}, and {Stillinger}}}]{SaDeSt98}
\bibinfo{author}{\bibfnamefont{S.}~\bibnamefont{{Sastry}}},
  \bibinfo{author}{\bibfnamefont{P.}~\bibnamefont{{Debenedetti}}},
  \bibnamefont{and}
  \bibinfo{author}{\bibfnamefont{F.}~\bibnamefont{{Stillinger}}},
  \bibinfo{journal}{Nature} \textbf{\bibinfo{volume}{393}},
  \bibinfo{pages}{554} (\bibinfo{year}{1998}).

\bibitem[{\citenamefont{Sillescu}(1999)}]{Sillescu99}
\bibinfo{author}{\bibfnamefont{H.}~\bibnamefont{Sillescu}},
  \bibinfo{journal}{J. of Non-Cryst. Sol.} \textbf{\bibinfo{volume}{243}},
  \bibinfo{pages}{81} (\bibinfo{year}{1999}).

\bibitem[{\citenamefont{Ediger}(2000)}]{Ediger00}
\bibinfo{author}{\bibfnamefont{M.~D.} \bibnamefont{Ediger}},
  \bibinfo{journal}{Annu. Rev. Phys. Chem.} \textbf{\bibinfo{volume}{51}},
  \bibinfo{pages}{99} (\bibinfo{year}{2000}).

\bibitem[{\citenamefont{Richert}(2002)}]{Richert02}
\bibinfo{author}{\bibfnamefont{R.}~\bibnamefont{Richert}}, \bibinfo{journal}{J.
  Phys.: Condens. Matter} \textbf{\bibinfo{volume}{14}}, \bibinfo{pages}{R703}
  (\bibinfo{year}{2002}).

\bibitem[{\citenamefont{Kob et~al.}(1997)\citenamefont{Kob, Donati, Plimpton,
  Poole, and Glotzer}}]{KoDoPlPoGl97}
\bibinfo{author}{\bibfnamefont{W.}~\bibnamefont{Kob}},
  \bibinfo{author}{\bibfnamefont{C.}~\bibnamefont{Donati}},
  \bibinfo{author}{\bibfnamefont{S.~J.} \bibnamefont{Plimpton}},
  \bibinfo{author}{\bibfnamefont{P.~H.} \bibnamefont{Poole}}, \bibnamefont{and}
  \bibinfo{author}{\bibfnamefont{S.~C.} \bibnamefont{Glotzer}},
  \bibinfo{journal}{Phys. Rev. Lett.} \textbf{\bibinfo{volume}{79}},
  \bibinfo{pages}{2827} (\bibinfo{year}{1997}).

\bibitem[{\citenamefont{Donati et~al.}(1998)\citenamefont{Donati, Douglas, Kob,
  Plimpton, Poole, and Glotzer}}]{DoDoKoPlPoGl98}
\bibinfo{author}{\bibfnamefont{C.}~\bibnamefont{Donati}},
  \bibinfo{author}{\bibfnamefont{J.}~\bibnamefont{Douglas}},
  \bibinfo{author}{\bibfnamefont{W.}~\bibnamefont{Kob}},
  \bibinfo{author}{\bibfnamefont{S.~J.} \bibnamefont{Plimpton}},
  \bibinfo{author}{\bibfnamefont{P.~H.} \bibnamefont{Poole}}, \bibnamefont{and}
  \bibinfo{author}{\bibfnamefont{S.~C.} \bibnamefont{Glotzer}},
  \bibinfo{journal}{Phys. Rev. Lett.} \textbf{\bibinfo{volume}{80}},
  \bibinfo{pages}{2338} (\bibinfo{year}{1998}).

\bibitem[{\citenamefont{Glotzer et~al.}(2000)\citenamefont{Glotzer, Novikov,
  and Schroder}}]{GlNoSc00}
\bibinfo{author}{\bibfnamefont{S.~C.} \bibnamefont{Glotzer}},
  \bibinfo{author}{\bibfnamefont{V.~N.} \bibnamefont{Novikov}},
  \bibnamefont{and} \bibinfo{author}{\bibfnamefont{T.~B.}
  \bibnamefont{Schroder}}, \bibinfo{journal}{J. Chem. Phys.}
  \textbf{\bibinfo{volume}{112}}, \bibinfo{pages}{509} (\bibinfo{year}{2000}).

\bibitem[{\citenamefont{Donati et~al.}(2002)\citenamefont{Donati, Franz,
  Parisi, and Glotzer}}]{DoFrPaGl02}
\bibinfo{author}{\bibfnamefont{C.}~\bibnamefont{Donati}},
  \bibinfo{author}{\bibfnamefont{S.}~\bibnamefont{Franz}},
  \bibinfo{author}{\bibfnamefont{G.}~\bibnamefont{Parisi}}, \bibnamefont{and}
  \bibinfo{author}{\bibfnamefont{S.~C.} \bibnamefont{Glotzer}},
  \bibinfo{journal}{J. of Non-Cryst. Sol.} \textbf{\bibinfo{volume}{307-310}},
  \bibinfo{pages}{215} (\bibinfo{year}{2002}).

\bibitem[{\citenamefont{Broderix et~al.}(2000)\citenamefont{Broderix,
  Bhattacharya, Cavagna, Zippelius, and Giardina}}]{BrBhCaZiGi00}
\bibinfo{author}{\bibfnamefont{K.}~\bibnamefont{Broderix}},
  \bibinfo{author}{\bibfnamefont{K.~K.} \bibnamefont{Bhattacharya}},
  \bibinfo{author}{\bibfnamefont{A.}~\bibnamefont{Cavagna}},
  \bibinfo{author}{\bibfnamefont{A.}~\bibnamefont{Zippelius}},
  \bibnamefont{and} \bibinfo{author}{\bibfnamefont{I.}~\bibnamefont{Giardina}},
  \bibinfo{journal}{Phys. Rev. Lett.} \textbf{\bibinfo{volume}{85}},
  \bibinfo{pages}{5360} (\bibinfo{year}{2000}).

\bibitem[{\citenamefont{Angelani et~al.}(2000)\citenamefont{Angelani, Leonardo,
  Ruocco, Scala, and Sciortino}}]{AnDLRuScSc00}
\bibinfo{author}{\bibfnamefont{L.}~\bibnamefont{Angelani}},
  \bibinfo{author}{\bibfnamefont{R.~D.} \bibnamefont{Leonardo}},
  \bibinfo{author}{\bibfnamefont{G.}~\bibnamefont{Ruocco}},
  \bibinfo{author}{\bibfnamefont{A.}~\bibnamefont{Scala}}, \bibnamefont{and}
  \bibinfo{author}{\bibfnamefont{F.}~\bibnamefont{Sciortino}},
  \bibinfo{journal}{Phys. Rev. Lett.} \textbf{\bibinfo{volume}{85}},
  \bibinfo{pages}{5356} (\bibinfo{year}{2000}).

\bibitem[{\citenamefont{Grigera et~al.}(2002)\citenamefont{Grigera, Cavagna,
  Giardina, and Parisi}}]{GrCaGiPa02}
\bibinfo{author}{\bibfnamefont{T.~S.} \bibnamefont{Grigera}},
  \bibinfo{author}{\bibfnamefont{A.}~\bibnamefont{Cavagna}},
  \bibinfo{author}{\bibfnamefont{I.}~\bibnamefont{Giardina}}, \bibnamefont{and}
  \bibinfo{author}{\bibfnamefont{G.}~\bibnamefont{Parisi}},
  \bibinfo{journal}{Phys. Rev. Lett.} \textbf{\bibinfo{volume}{88}},
  \bibinfo{pages}{55502} (\bibinfo{year}{2002}).

\bibitem[{\citenamefont{B\"uchner and Heuer}(2000)}]{BuHe00}
\bibinfo{author}{\bibfnamefont{S.}~\bibnamefont{B\"uchner}} \bibnamefont{and}
  \bibinfo{author}{\bibfnamefont{A.}~\bibnamefont{Heuer}},
  \bibinfo{journal}{Phys. Rev. Lett.} \textbf{\bibinfo{volume}{84}},
  \bibinfo{pages}{2168} (\bibinfo{year}{2000}).

\bibitem[{\citenamefont{{Glotzer} et~al.}(1998)\citenamefont{{Glotzer}, {Jan},
  and {Poole}}}]{GlJaLoMaPo98}
\bibinfo{author}{\bibfnamefont{S.}~\bibnamefont{{Glotzer}}},
  \bibinfo{author}{\bibfnamefont{N.}~\bibnamefont{{Jan}}}, \bibnamefont{and}
  \bibinfo{author}{\bibfnamefont{P.}~\bibnamefont{{Poole}}},
  \bibinfo{journal}{Phys. Rev. E} \textbf{\bibinfo{volume}{57}},
  \bibinfo{pages}{7350} (\bibinfo{year}{1998}).

\bibitem[{\citenamefont{Barrat and Zecchina}(1999)}]{BaZe99}
\bibinfo{author}{\bibfnamefont{A.}~\bibnamefont{Barrat}} \bibnamefont{and}
  \bibinfo{author}{\bibfnamefont{R.}~\bibnamefont{Zecchina}},
  \bibinfo{journal}{Phys. Rev. E} \textbf{\bibinfo{volume}{59}},
  \bibinfo{pages}{R1299} (\bibinfo{year}{1999}).

\bibitem[{\citenamefont{{Poole} et~al.}(1997)\citenamefont{{Poole}, {Coniglio},
  {Glotzer}, and {Jan}}}]{PoCoGlJa97}
\bibinfo{author}{\bibfnamefont{P.}~\bibnamefont{{Poole}}},
  \bibinfo{author}{\bibfnamefont{A.}~\bibnamefont{{Coniglio}}},
  \bibinfo{author}{\bibfnamefont{S.}~\bibnamefont{{Glotzer}}},
  \bibnamefont{and} \bibinfo{author}{\bibfnamefont{N.}~\bibnamefont{{Jan}}},
  \bibinfo{journal}{Phys. Rev. Lett.} \textbf{\bibinfo{volume}{78}},
  \bibinfo{pages}{3394} (\bibinfo{year}{1997}).

\bibitem[{\citenamefont{Nicodemi and Coniglio}(1997)}]{NiCo97}
\bibinfo{author}{\bibfnamefont{M.}~\bibnamefont{Nicodemi}} \bibnamefont{and}
  \bibinfo{author}{\bibfnamefont{A.}~\bibnamefont{Coniglio}},
  \bibinfo{journal}{J. Phys. A} \textbf{\bibinfo{volume}{30}},
  \bibinfo{pages}{L187} (\bibinfo{year}{1997}).

\bibitem[{\citenamefont{Lawlor et~al.}(2002)\citenamefont{Lawlor, Reagan,
  McCullagh, Gregorio, and Dawson}}]{LaReMcGrTaDa02}
\bibinfo{author}{\bibfnamefont{A.}~\bibnamefont{Lawlor}},
  \bibinfo{author}{\bibfnamefont{D.}~\bibnamefont{Reagan}},
  \bibinfo{author}{\bibfnamefont{G.~D.} \bibnamefont{McCullagh}},
  \bibinfo{author}{\bibfnamefont{P.~D.} \bibnamefont{Gregorio}},
  \bibnamefont{and} \bibinfo{author}{\bibfnamefont{P.~T.~K.}
  \bibnamefont{Dawson}}, \bibinfo{journal}{Phys. Rev. Lett.}
  \textbf{\bibinfo{volume}{89}}, \bibinfo{pages}{245503}
  (\bibinfo{year}{2002}).

\bibitem[{\citenamefont{Dawson et~al.}(2002)\citenamefont{Dawson, Lawlor,
  De~Gregorio, McCullagh, Zaccarelli, and Tartaglia}}]{DaLaGrMcZaTa02}
\bibinfo{author}{\bibfnamefont{K.~A.} \bibnamefont{Dawson}},
  \bibinfo{author}{\bibfnamefont{A.}~\bibnamefont{Lawlor}},
  \bibinfo{author}{\bibfnamefont{P.}~\bibnamefont{De~Gregorio}},
  \bibinfo{author}{\bibfnamefont{G.~D.} \bibnamefont{McCullagh}},
  \bibinfo{author}{\bibfnamefont{E.}~\bibnamefont{Zaccarelli}},
  \bibnamefont{and}
  \bibinfo{author}{\bibfnamefont{P.}~\bibnamefont{Tartaglia}},
  \bibinfo{journal}{Physica A} \textbf{\bibinfo{volume}{316}},
  \bibinfo{pages}{115} (\bibinfo{year}{2002}).

\bibitem[{\citenamefont{Fierro et~al.}(2000)\citenamefont{Fierro, {de Candia},
  and Coniglio}}]{FiCaCo00}
\bibinfo{author}{\bibfnamefont{A.}~\bibnamefont{Fierro}},
  \bibinfo{author}{\bibfnamefont{A.}~\bibnamefont{{de Candia}}},
  \bibnamefont{and} \bibinfo{author}{\bibfnamefont{A.}~\bibnamefont{Coniglio}},
  \bibinfo{journal}{Phys. Rev. E} \textbf{\bibinfo{volume}{62}},
  \bibinfo{pages}{7715} (\bibinfo{year}{2000}).

\bibitem[{\citenamefont{Stariolo and Arenzon}(1999)}]{StAr99}
\bibinfo{author}{\bibfnamefont{D.~A.} \bibnamefont{Stariolo}} \bibnamefont{and}
  \bibinfo{author}{\bibfnamefont{J.~J.} \bibnamefont{Arenzon}},
  \bibinfo{journal}{Phys. Rev. E} \textbf{\bibinfo{volume}{59}},
  \bibinfo{pages}{R4762} (\bibinfo{year}{1999}).

\bibitem[{\citenamefont{Arenzon et~al.}(2000)\citenamefont{Arenzon,
  Ricci-Tersenghi, and Stariolo}}]{ArRiSt00}
\bibinfo{author}{\bibfnamefont{J.~J.} \bibnamefont{Arenzon}},
  \bibinfo{author}{\bibfnamefont{F.}~\bibnamefont{Ricci-Tersenghi}},
  \bibnamefont{and} \bibinfo{author}{\bibfnamefont{D.~A.}
  \bibnamefont{Stariolo}}, \bibinfo{journal}{Phys. Rev. E}
  \textbf{\bibinfo{volume}{62}}, \bibinfo{pages}{5978} (\bibinfo{year}{2000}).

\bibitem[{\citenamefont{Arenzon et~al.}(1996)\citenamefont{Arenzon, Nicodemi,
  and Sellitto}}]{ArNiSe96}
\bibinfo{author}{\bibfnamefont{J.~J.} \bibnamefont{Arenzon}},
  \bibinfo{author}{\bibfnamefont{M.}~\bibnamefont{Nicodemi}}, \bibnamefont{and}
  \bibinfo{author}{\bibfnamefont{M.}~\bibnamefont{Sellitto}},
  \bibinfo{journal}{J. Physique I} \textbf{\bibinfo{volume}{6}},
  \bibinfo{pages}{1143} (\bibinfo{year}{1996}).

\bibitem[{\citenamefont{Crisanti and Leuzzi}(2002)}]{CrLe02}
\bibinfo{author}{\bibfnamefont{A.}~\bibnamefont{Crisanti}} \bibnamefont{and}
  \bibinfo{author}{\bibfnamefont{L.}~\bibnamefont{Leuzzi}},
  \bibinfo{journal}{Phys. Rev. Lett.} \textbf{\bibinfo{volume}{89}},
  \bibinfo{pages}{237204} (\bibinfo{year}{2002}).

\bibitem[{\citenamefont{de~Candia and Coniglio}(2001)}]{CaCo01}
\bibinfo{author}{\bibfnamefont{A.}~\bibnamefont{de~Candia}} \bibnamefont{and}
  \bibinfo{author}{\bibfnamefont{A.}~\bibnamefont{Coniglio}},
  \bibinfo{journal}{Phys. Rev. E} \textbf{\bibinfo{volume}{65}},
  \bibinfo{pages}{016132} (\bibinfo{year}{2001}).

\bibitem[{\citenamefont{Klein et~al.}(1994)\citenamefont{Klein, Gould, Ramos,
  Clejan, and Mel'cuk}}]{KlGoRaClMe94}
\bibinfo{author}{\bibfnamefont{W.}~\bibnamefont{Klein}},
  \bibinfo{author}{\bibfnamefont{H.}~\bibnamefont{Gould}},
  \bibinfo{author}{\bibfnamefont{R.~A.} \bibnamefont{Ramos}},
  \bibinfo{author}{\bibfnamefont{I.}~\bibnamefont{Clejan}}, \bibnamefont{and}
  \bibinfo{author}{\bibfnamefont{A.~I.} \bibnamefont{Mel'cuk}},
  \bibinfo{journal}{Physica A} \textbf{\bibinfo{volume}{205}},
  \bibinfo{pages}{738} (\bibinfo{year}{1994}).

\bibitem[{\citenamefont{Johnson et~al.}(1998)\citenamefont{Johnson, Mel'cuk,
  Gould, Klein, and Mountain}}]{JoMeGoKlMo98}
\bibinfo{author}{\bibfnamefont{G.}~\bibnamefont{Johnson}},
  \bibinfo{author}{\bibfnamefont{A.~I.} \bibnamefont{Mel'cuk}},
  \bibinfo{author}{\bibfnamefont{H.}~\bibnamefont{Gould}},
  \bibinfo{author}{\bibfnamefont{W.}~\bibnamefont{Klein}}, \bibnamefont{and}
  \bibinfo{author}{\bibfnamefont{R.~D.} \bibnamefont{Mountain}},
  \bibinfo{journal}{Phys. Rev. E} \textbf{\bibinfo{volume}{57}},
  \bibinfo{pages}{5707} (\bibinfo{year}{1998}).

\bibitem[{\citenamefont{Castillo et~al.}(2002)\citenamefont{Castillo, Chamon,
  Cugliandolo, and Kennett}}]{CaChCuKe02}
\bibinfo{author}{\bibfnamefont{H.~E.} \bibnamefont{Castillo}},
  \bibinfo{author}{\bibfnamefont{C.}~\bibnamefont{Chamon}},
  \bibinfo{author}{\bibfnamefont{L.~F.} \bibnamefont{Cugliandolo}},
  \bibnamefont{and} \bibinfo{author}{\bibfnamefont{M.~P.}
  \bibnamefont{Kennett}}, \bibinfo{journal}{Phys. Rev. Lett.}
  \textbf{\bibinfo{volume}{88}}, \bibinfo{pages}{237201}
  (\bibinfo{year}{2002}).

\bibitem[{\citenamefont{Montanari and Ricci-Tersenghi}(2003)}]{MoRi03}
\bibinfo{author}{\bibfnamefont{A.}~\bibnamefont{Montanari}} \bibnamefont{and}
  \bibinfo{author}{\bibfnamefont{F.}~\bibnamefont{Ricci-Tersenghi}},
  \bibinfo{journal}{Phys. Rev. Lett.} \textbf{\bibinfo{volume}{90}},
  \bibinfo{pages}{017203} (\bibinfo{year}{2003}).

\bibitem[{\citenamefont{Berthier}(2003)}]{Berthier03b}
\bibinfo{author}{\bibfnamefont{L.}~\bibnamefont{Berthier}}
  (\bibinfo{year}{2003}), \bibinfo{note}{cond-mat/0303453}.

\bibitem[{\citenamefont{Oshanin et~al.}(2003)\citenamefont{Oshanin, B\'enichou,
  Burlatsky, and Moreau}}]{OsBeBuMo02}
\bibinfo{author}{\bibfnamefont{G.}~\bibnamefont{Oshanin}},
  \bibinfo{author}{\bibfnamefont{O.}~\bibnamefont{B\'enichou}},
  \bibinfo{author}{\bibfnamefont{S.~F.} \bibnamefont{Burlatsky}},
  \bibnamefont{and} \bibinfo{author}{\bibfnamefont{M.}~\bibnamefont{Moreau}},
  in \emph{\bibinfo{booktitle}{Instabilities and {N}on-{E}quilibrium
  {S}tructures {IX}}}, edited by
  \bibinfo{editor}{\bibfnamefont{E.}~\bibnamefont{Tirapegui}} \bibnamefont{and}
  \bibinfo{editor}{\bibfnamefont{O.}~\bibnamefont{Descalzi}}
  (\bibinfo{publisher}{Kluwer Academic Pub.}, \bibinfo{year}{2003}),
  \bibinfo{note}{to appear}.

\end{thebibliography}

\end{document}